\documentclass[smallcondensed]{svjour3}    

\smartqed  
\usepackage{graphicx}

\begin{document}

\title{Trapping in three-planet resonances during gas-driven migration}

\author{Anne-Sophie Libert \and Kleomenis Tsiganis}

\institute{A.-S. Libert \at
              naXys, Department of Mathematics, University of Namur, 8 Rempart de la Vierge, B5000 Namur, Belgium \\
              \email{anne-sophie.libert@fundp.ac.be}
	      \and K. Tsiganis \at Department of Physics, University of Thessaloniki, GR-54 124 Thessaloniki, Greece \\
	      \email{tsiganis@astro.auth.gr}}        
\maketitle

\begin{abstract}

We study the establishment of three-planet resonances -- similar to the {\it Laplace} resonance in the 
Galilean satellites -- and their effects on the mutual inclinations of the orbital planes of the planets, assuming 
that the latter undergo migration in a gaseous disc. In particular, we examine the resonance relations that occur, 
by varying the physical and initial orbital parameters of the planets (mass, initial semi-major axis and eccentricity) 
as well as the parameters of the migration forces (migration rate and eccentricity damping rate), which are modeled 
here through a simplified analytic prescription. We find that, in general, for planetary masses 
 below $1.5~M_{\rm J}$, multiple-planet resonances of the form $n_3$:$n_2$:$n_1$=1:2:4 and 1:3:6 are established, as the inner planets, $m_1$ and $m_2$, get trapped in a 1:2 
resonance and the outer planet $m_3$ subsequently is captured in a 1:2 or 1:3 resonance with $m_2$. For mild eccentricity damping, 
the resonance pumps the eccentricities of all planets on a relatively short time-scale, to the point where they 
enter an inclination-type resonance (as in Libert \& Tsiganis 2011); then mutual inclinations can grow to $\sim 
35^{\circ}$, thus forming a {\it ``3-D system''}. On the other hand, we find that trapping of $m_2$ in a 2:3 
resonance with $m_1$ occurs very rarely, for the range of masses used here, so only two cases of capture in a 
respective three-planet resonance  were found. Our results suggest that trapping in a three-planet resonance 
can be common in exoplanetary systems, provided that the planets are not very massive. Inclination pumping could then occur relatively fast, provided that eccentricity damping is not very efficient so that at least one of the inner 
planets acquires an orbital eccentricity higher than $e=0.3$. 


\end{abstract}

\section{Introduction}
\label{intro}

Recently, the possibility that extrasolar planetary systems can be '3-D systems', namely they are composed of two or 
more planets whose orbital planes have substantial values of mutual inclinations, has been considered. Some analytical studies have highlighted that such systems can be long-term stable, either following normal secular dynamics or due to the action 
of a phase-protection mechanism, such as a mean-motion resonance (MMR) or a secular, Kozai-type resonance (e.g.\ 
Michtchenko et al. 2006b, Libert \& Henrard 2007, 2008, Libert \& Tsiganis 2009a). A first observational confirmation 
(McArthur et al.\ 2010) for the $\upsilon$ Andromedae system estimated the mutual inclination of the orbital planes 
of planets $c$ and $d$ of this system to $\sim30^\circ$. 

In the meantime, a number of studies on the possible formation mechanisms of 3-D planetary systems have been realized. Strong dynamical instability of systems with giant planets (planet-planet scattering) is generally invoked to explain the orbital properties of extrasolar systems, in particular the highly eccentric orbits seen in many (see e.g.\ Ford, Lystad \& Rasio 2005, Fabrycky \& Tremaine 2007, Ford \& Rasio 2008, Nagasawa, Ida \& Bessho 2008). Many of these have shown that the most common outcome of gravitational scattering by close encounters is the hyperbolic ejection of one planet, the 'survivors' having significant values of eccentricity and mutual inclination (e.g.\ Marzari \& Weidenschilling 2002, Chatterjee et al.\ 2008, Juri\'c \& Tremaine 2008). These works focused on gas-free systems, assuming that systems that are relatively compact when the gas nebula dissipates will 
undergo planet-planet scattering after a relatively short time. Only a few works investigated the combined action of disc torques and planet-planet scattering (e.g.\ Adams \& Laughlin 2003, Moorhead \& Adams 2005, Thommes, Matsumura \& Rasio 2008, Matsumura et al.\ 2010). However, an unstable crowded system -- the starting point of typical planet-planet 
scattering simulations -- is not the unique result of formation in a gas-disc, since resonant interactions during 
the gas phase should lead to stable resonant systems, such as the ones observed (e.g. Michtchenko et al. 2006a, Hadjidemetriou 2006, 2009 for studies of mean-motion resonant systems). 

Resonant two-planet systems can also be `3-D', as first shown by Thommes \& Lissauer (2003), who studied the evolution of planets trapped in a 1:2 MMR, under the effects of gas-driven (Type II) migration, inside the protoplanetary gas disc. Libert \& Tsiganis (2009b) extended the study of this 'inclination-type resonance' to capture in higher order resonances (such as the 2:5, 
1:3, 1:4 and 1:5), showing that inclination-excitation occurs, as long as the eccentricity of one planet exceeds $\sim 0.4$ and the inner planet is not very heavy. 

Concerning resonant systems with more than two planets, it has been recently shown that, at least in the case of low-mass planets such as the ones in our solar system, gas-driven migration can force the planets to enter into a {\it multiple-planet resonance}, an analogue of the Laplace resonance in the Galilean moons (Morbidelli et al.\ 2007). At present, at least three extrasolar systems are believed to be in multiple-planet resonance: HR8799 (e.g. Reidemeister et al. 2009) and Gl876 (Rivera et al. 2010) in a Laplace configuration, and the KOI500 five-planet exosystem that shows two three-body resonances (Lissauer et al. 2011). Moreover, simulations on a possible trapping of HD82943 exosystem in a Laplace resonance were also performed by Beaug\'e et al. (2008). Libert \& Tsiganis (2011) showed that a Laplace three-planet resonance can be achieved for Jupiter-mass planets, but generally becomes unstable as the resonance excites the eccentricities of all planets and planet-planet scattering sets in. This mechanism, combining the action of disc torques and planet-planet scattering, typically results in the ejection of one of the three planets, leaving behind a two-planet system in a nearly hierarchical configuration and with median mutual inclination of $\sim 30^\circ$. In $10\%$ of the simulations presented in that paper, the triple resonance remains stable for sufficiently long times, leading to the formation of a `3-D' system, through resonant inclination excitation, similarly to the two-body inclination-type resonance mentioned before. 

As also noted in Morbidelli et al.\ (2007), a multiple-planet resonance is a delicate dynamical configuration; not all resonant ratios can be reached by all planetary masses (or mass ratios) and not all resonances are long-term stable. In Libert \& Tsiganis (2011), we only investigated the dynamical disruption of the Laplace 1:2:4 configuration. The goal of the present contribution 
is to examine for what masses of the planets and parameters of the disc a triple resonance can be established, and of what type. Also, we examine the frequency of inclination excitation, in three-planet resonances. As expected the initial orbital separation of the three planets plays a major role in the final outcome and this issue is adressed also in the following. 

The paper is organized as follows. In Section 2, we describe the set-up of our numerical experiments, while the results of these simulations are presented in detail in Section 3. The eccentricity distribution leading to inclination excitation is presented in Section 4. Finally, our conclusions are given in Section 5.

\section{Numerical model}

A commonly accepted scenario for the origin of resonant configurations assumes disc-induced differential orbital migration of the planets, initially much further apart, towards their parent star. While a Jupiter-sized planet carves a gap in the gaseous disc, repelling material away from its neighborhood, the massive gaseous disc tends to refill the gap, by viscous diffusion. The balance between these processes leads to a relatively slow inward migration of the planets, called Type II migration. An order-of-magnitude estimate for the migration rate of a planet was given by Ward (1997):
\begin{equation}\label{rate}
\left|\frac{\dot{a}}{a}\right| \sim 9.4 \alpha \left(\frac{H}{a} \right)^2 
P^{-1}
\end{equation}
where $P$ is the orbital period of the body. In the following, the Shakura-Sunyaev viscosity parameter is set to $\alpha=4\times 10^{-3}$ 
(Shakura \& Sunyaev 1973) and the aspect ratio of the disc to $H/a=0.05$. Equation (1) is used to derive an estimate for the constant drift rate that we impose on the planet, as in Libert \& Tsiganis (2009b): we assume $a$ in Equation (\ref{rate}) to be constant and equal to the initial value of semi-major axis of the planet's orbit. This is of course a simplified prescription, but adequate for analyzing the dynamics of resonance encounters. As planet-disc interactions also affect the eccentricity of the migrating planet (see Goldreich \& Tremaine 1980, Papaloizou, Nelson
\& Masset 2001), exponential damping is also assumed in the following way:
\begin{equation}\label{damprate}
\frac{\dot{e}}{e}=-K\left|\frac{\dot{a}}{a}\right|,
\end{equation}
\noindent
where $K$ is chosen to be 0, 1 and 5 in the following. 

We start our simulations with a system of three planets ($m_1$, $m_2$ and $m_3$; $m_1$ being the inner body and $m_3$ the outer one) evolving around a 1~$M_{\odot}$ star. Assuming that the inner disc is likely to be largely depleted by the time our planets were formed, we simulate planet-disc interaction by applying a suitable Stokes-type drag force  (following Beaug\'e, Michtchenko \& Ferraz-Mello 2005) on $m_2$ and $m_3$ only. To integrate this four-body problem we use a variant of the SYMBA code (Duncan et al.\ 1998), which is able to deal symplectically with close encounters between massive bodies, in which the drag terms are added in the equations of motion, introducing an exponential drift in semi-major axis (migration) and eccentricity (damping). Let us note that, even if the drift is not applied on the inner body, when capture in resonance occurs, the two bodies subsequently migrate as a pair towards the star -- at a slower rate -- and both their eccentricities are damped. However, for mild eccentricity damping, the growth of the eccentricities may be very rapid, as migration in resonance continues (e.g.\ Lee \& Peale 2002, Ferraz-Mello et al.\ 2003). Note that, as shown in Morbidelli \& Crida (2007), in more realistic simulations of gas-planet interactions the resonance pair may stop migrating (or even start migrating outwards), depending on the planetary masses, $\alpha$ and $H/a$. However, in such a case, $m_3$ would approach a resonance with $m_2$ more easily and a triple resonance could again be established. To take into account (partly) these uncertainties that are related to our simplified migration model, we perform our simulations also using migration rates 5 times larger and 5 times smaller than 
the values given by Equation\ (1).   

In Libert \& Tsiganis (2011), the establishment of the Laplace triple resonance was reached by a two-step formation process, following the lines of Morbidelli et al.\ (2007): first we simulated the 1:2 resonant capture of the two inner planets, $m_1$ and $m_2$. Then, planet $m_3$ was introduced in the simulation and forced to migrate into the 1:2 MMR with $m_2$. The Laplace 
resonance was thus approached, but not always established; the orbits of heavy giant planets usually become chaotic and planet-planet scattering dissolves the system before the establishment of the 1:2:4 resonance. Here, the methodology is different, as 
we do not focus on this particular resonance. Instead, we try to identify which triple resonances can be established by disc-induced differential orbital migration of three planets.

Two initial configurations of the three-planet systems are examined in this work, depending on the position of the intermediate planet ($m_2$) with respect to its 1:2 MMR with $m_1$ (initially located at $1$). The semi-major axis of $m_2$ is set to $1.9$ ({\it exterior} to their 1:2 MMR - see Section \ref{firstconf}) and $1.4$ ({\it interior} to their 1:2 MMR - see Section \ref{secondconf}). In either case, we consider several values of the semi-major axis of $m_3$: $2.8$, $3.1$ and $3.6$ for the exterior case, $2$, $2.4$ and $2.7$ for the interior case. The initial eccentricities and inclinations of all planets are nearly zero ($e=0.001$, $i=0.01^\circ$). The effects of initially eccentric orbits are studied in Section \ref{var}. Nine mass configurations are considered for the planets: $m_1$ is fixed to $1.5$ $M_{J}$ ($M_J=$Jupiter's mass), while the mass of the second planet is set to $m_2=0.75$, $1.5$ or $3$ $M_{Jup}$ and $m_3$ to $m_2/2$, $m_2$ or $2m_2$. Our system of units is such that $G=1$ and $M_\odot=1$. Taking the semi-major axis of $m_1$ as the unit of distance ($a_1=1$), the time unit is defined such that the period of $m_1$ is $P_1=2 \pi$. All simulations spanned a time equal to $10^6$ time units, a time long enough to be comparable to the disc's lifetime. The results of these simulations are described in the following section.


\section{Results}

\subsection{Exterior configuration}\label{firstconf}

We consider first the evolution in the gas-disc of a three-planet system in which the intermediate planet ($m_2$) is initially set exterior to the location of its 1:2 MMR with $m_1$: here we use $a_1=1$ and $a_2=1.9$. Three different initial positions of the outer planet ($m_3$) are used in the following simulations: $2.8$ (initially outside the 2:3 MMR with $m_2$), $3.1$ (initially outside the 1:2 MMR with $m_2$) and $3.6$ (initially outside the 2:5 MMR with $m_2$). For the values of the mass ratios and eccentricity damping discussed in the previous section, we aim to determine the frequency of (i) capture in different multiple-planet resonance - if capture occurred - and (ii) subsequent resonant inclination excitation.

\begin{table}
\centering
\caption{MMR captures for three-planet systems of different masses ($m_1=1.5$ $M_{J}$), for different values of the eccentricity damping rate. The symbol '$*$' indicates that the inclination-type resonance occurs as the planets migrate in resonance. The migration rate is equal to the value given by Equation (\ref{rate}), namely $\left|\dot{a_2}/a_2\right|=5.71\times10^{-6}$ time units$^{-1}$ and $\left|\dot{a_3}/a_3\right|=3.19\times10^{-6}$, $2.74\times10^{-6}$ or $2.19\times10^{-6}$ time units$^{-1}$ for $a_3=2.8$, $3.1$ and $3.6$ respectively.}
\label{tableoutside}
\begin{tabular}{lllllllllll}
\hline
& \multicolumn{3}{r}{$a_3=2.8$} & \multicolumn{3}{r}{$3.1$}& \multicolumn{3}{r}{$3.6$}\\
$m_2$ & $m_3$ & K=0 & K=1 & K=5 & K=0 & K=1 & K=5 & K=0 & K=1 & K=5 \\
\hline
3  & 6 &  &  &  &  & 1:2 & 1:2 & 1:2 & 1:2 $*$ & 1:2 $*$ \\
3  & 3 &  &  &  & 1:2 & 1:2 & 1:2 & 1:2 $*$ & 1:2 $*$ & 1:2\\
3  & 1.5 &  &  & 1:2:4 & 1:2 $*$ & 1:2 & 1:2 $*$ & 1:2 $*$ & 1:4:8 $*$ & 1:2\\
1.5  & 3 &  & 1:2:4 &  & 1:2 & 1:2 & 1:2 & 1:3:6 $*$ & 1:3:6 $*$ & 1:3:6 \\
1.5  & 1.5 &  & 1:2:4 $*$ & 1:2:4 & 1:2 & 1:2 & 1:2 & 1:3:6 $*$ & 1:3:6 $*$ & 1:3:6\\
1.5  & 0.75 & 1:2:4 $*$ & 1:2:4 $*$ & 1:2:4 & 1:2 & 1:2 & 1:2:4 $*$ & 1:3:6 $*$ & 1:3:6 $*$ & 1:3:6\\
0.75  & 1.5 & 1:2:4 & 1:2:4 & 1:2:4 &  & 1:2 & 1:2:4 & 1:3:6 & 1:2 & 1:3:6 \\
0.75  & 0.75 & 1:2:4 & 1:2:4 $*$ & 1:2:4 &  &  & 1:2:4 & 1:3:6 $*$ & 1:2 & 1:2\\
0.75  & 0.375 & 1:2:4 $*$ & 1:2:4 $*$ & 1:2:4 &  & 1:2:4 &  & 1:2 $*$ & 1:2 $*$ & 1:2 \\
\hline
\end{tabular}
\end{table}

The results of these simulations are presented in Table \ref{tableoutside}. For each mass ratio and eccentricity damping value, we describe the time evolution of the system by denoting the MMR in which it is captured. Three different outcomes are observed in the simulations:

\noindent (a) the unstable case, represented by an empty slot in the Table. The system is destabilized in less than $10^5$ time units, 
without having previously reached any resonant configuration\footnote{We note that the simulation is stopped if one of the planets is ejected from the system or two planets merge.}. As seen in the Table, this occurs primarily for mass ratios $m_2/m_1\geq 1$ and 
$m_3/m_1 \geq 1$, i.e.\ for more massive outer planets, and small orbital separations. As $K$ increases, the instability seems to occur only for $m_2/m_1 \geq 2$ and $m_3/m_1 \geq 2$. This implies that, for a massive system in a compact initial configuration, the eccentricities grow fast and the planetary orbits cross each other, before a resonance can be established. \\
(b) the two-body resonant case. The two inner planets are trapped in a 1:2 resonant configuration, which is found to be destabilized when the outer planet is approaching a resonance with $m_2$. Note that this occurs for an initial $a_3=3.1$ or higher. For an initial $a_3=2.8$, no similar capture of the two inner planets in a 2:3 MMR is observed. \\
(c) the three-body resonant case, where all three planets are trapped in a resonant configuration. According to the Table, there are two main resonances, depending on the initial configuration (i.e.\ period ratio $P_3/P_2$). Inclination excitation is observed for both resonances, but only for $K\leq 1$. This implies that the excitation is related to eccentricity growth in the resonance, as was also shown in Libert \& Tsiganis (2011).

As Table \ref{tableoutside} shows, massive planets (first lines of Table \ref{tableoutside}) and an initial location of the intermediate planet very close to its 1:2 MMR with the outer planet ($a_3=3.1$ - columns 6, 7 and 8 of Table \ref{tableoutside}) prevent the formation of a multi-resonant state. For massive planets, the typical outcome is the ejection of one of the bodies from the system at the beginning of the simulation, especially for the particularly compact initial configuration $a_1=1$, $a_2=1.9$ and $a_3=2.8$. For the two other initial configurations, i.e.\ larger distance of $m_3$ from the inner planets, massive planets meet easily the 1:2 MMR between $m_1$ and $m_2$, but no capture occurs for $m_3$. These two configurations excepted (massive planets and $a_3=3.1$), most of the simulations result in three-planet resonant configurations. The two dominant relations found are: the Laplace 1:2:4 resonance (for an initial $a_3=2.8$) and the 1:3:6 resonance (for an initial $a_3>3$). The notation used here means that $m_1$ and $m_2$ are in a 1:2 MMR ($n_2$:$n_1$=1:2), while $m_2$ and $m_3$ in a 1:3 MMR ($n_3$:$n_2$=1:3). Thus, the multiple-planet resonance is labeled as $n_3$:$n_2$:$n_1$=1:3:6. The same convention will always be adopted in the following.

Note that these results may depend on the assumed simplified migration model, as well as on the assumed migration rates (see Section \ref{var}). However, we believe that the observed trend, i.e.\ easier establishment of a triple resonance for less massive triplets should hold, unless the relative migration of planets is strongly affected by variations in the surface density profile of the gas disc (e.g.\ as in Morbidelli \& Crida 2007).

\begin{figure}
\centering
\rotatebox{270}{\includegraphics[height=12cm]{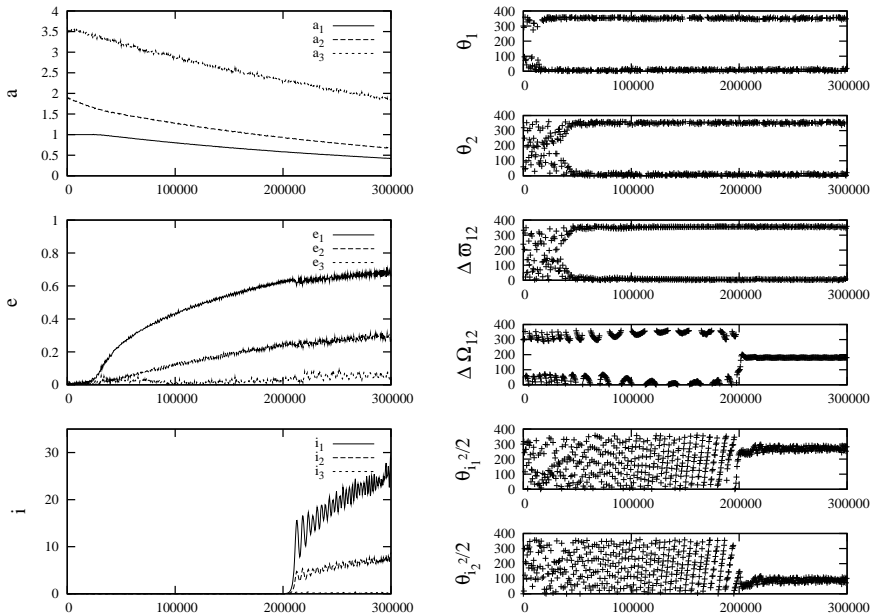}}
\caption{Inclination excitation for a 1:2 MMR between $m_1$ and $m_2$. The planetary masses are $m_2=3$ and $m_3=1.5 M_{J}$. Initial location of the outer planet is $a_3=3.6$. No eccentricity damping is considered ($K=0$).}
\label{figthommes}
\end{figure}

\begin{figure}
\centering
\rotatebox{270}{\includegraphics[height=12cm]{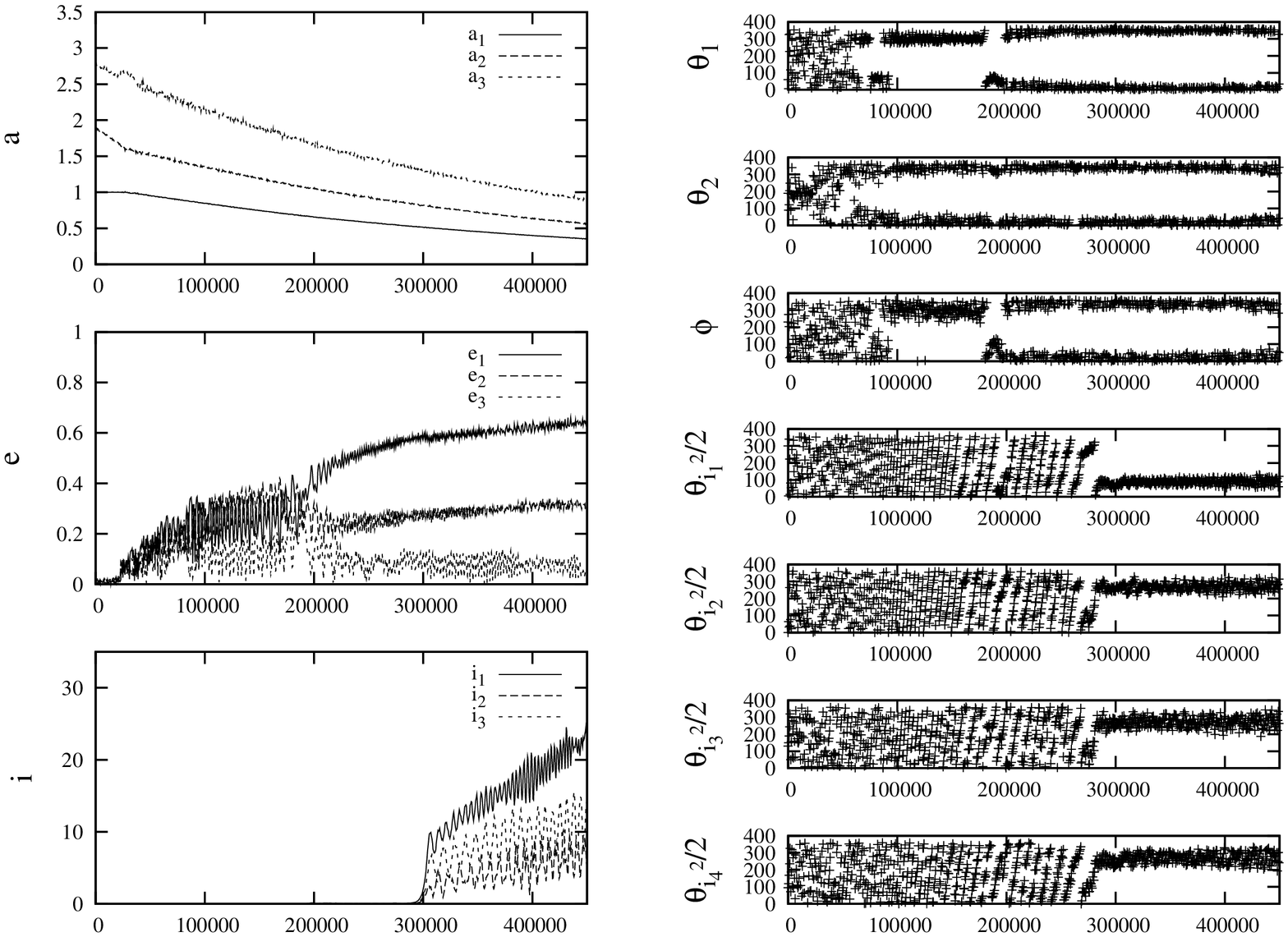}}
\caption{Inclination excitation for a three-planet system in a Laplace 1:2:4 resonant configuration. The planetary masses are $m_2=1.5, m_3=0.75 M_{J}$. Initial location of the outer planet is $a_3=2.8$ and eccentricity damping is set to $K=1$.}
\label{figlaplace}
\end{figure}

As the system remains in MMR and continues to migrate, the eccentricities of the planets become high enough that the system enters an inclination-type resonance, which induces rapid growth of the inclinations. Inclination excitation is indicated by the presence of the symbol '$*$' in Table \ref{tableoutside}. Inclination excitation can arise from two resonant mechanisms: the one described by Thommes \& Lissauer (2003) for two migrating planets in the 1:2 MMR and a similar one acting with three planets in a triply-resonant configuration, described in Libert \& Tsiganis (2011). Examples of both outcomes are shown in Figures \ref{figthommes} and \ref{figlaplace} respectively.

Figure \ref{figthommes} shows the evolution of a three-planet system where two planets go through the Thommes \& Lissauer's mechanism. While the outer planets migrate, $m_1$ and $m_2$ are captured in a 1:2 MMR at about $3\times10^{4}$ time units, characterized by the libration of the resonant angles $\theta_{1}=\lambda_{1}-2\lambda_{2}+\varpi_{1}$ and  $\theta_{2}=\lambda_{1}-2\lambda_{2}+\varpi_{2}$ around $0^\circ$. As a result, the difference of the longitudes of pericenter $\Delta \varpi=\varpi_{1}-\varpi_{2}=\theta_1-\theta_2$ starts to oscillate around $0^\circ$, i.e.\ the planets are in apsidal alignment. As both planets continue to migrate while in resonance, their eccentricities increase. When their values are high enough, the system enters an inclination-type resonance: the angles $\theta_{i_{1}^2}= 2\lambda_{1}-4\lambda_{2}+2\Omega_{1}$ and $\theta_{i_{2}^2}=2\lambda_{1}-4\lambda_{2}+2\Omega_{2}$ start to librate at about $2\times 10^{5}$~time units. A rapid growth of the inclinations of both planets is then observed, as well as a libration of the relative longitude of the nodes, $\Delta \Omega=\Omega_{1}-\Omega_{2}=(\theta_{i_{1}^2}-\theta_{i_{2}^2})/2$, around the anti-alignment state. Note that the eccentricity and inclination of the outer planet $m_3$, which is not in resonance with the other ones, remain almost constant (i.e.\ nearly zero) during the whole simulation of Figure \ref{figthommes}. The body $m_3$ goes successively through 1:3 (when $a_2\simeq1.6$ and $a_3\simeq3.3$) and 1:4 (when $a_2\simeq0.93$ and $a_3\simeq2.3$) commensurabilities with $m_2$, and at $\sim 3.5 \times 10^5$ time units (not shown on the figure), while approaching 1:5 commensurability ($a_2\simeq0.6$ and $a_3\simeq1.76$), destabilizes the whole system, leading to planet-planet scattering and the subsequent formation of a non-coplanar two-planet system, the less massive planet being ejected from the system. Indeed, the given $\dot{a}_2$ and masses result in a given change in energy of the resonant pair which could be translated to a mean migration rate. If this rate is smaller than $\dot{a}_3$, then we have divergent migration of $m_3$. A further analysis of the reason for instability in all studied systems (i.e. MMR crossing during divergent migration or instability before the resonance) is reserved for future work. Let us note that for the same initial configuration of the system of Figure \ref{figthommes} but for less massive planets, a capture in a multiple-resonance generally happens, as shown in Table \ref{tableoutside} (lines 4 to 9).

Inclination excitation due to multiple-planet resonance is described in Figure\ \ref{figlaplace}. While the outer planet migrates, the two inner planets are captured in a 1:2 MMR at about $3\times10^{4}$ time units, characterized by the libration of both resonant angles $\theta_1=\lambda_1-2\lambda_2+\varpi_1$ and $\theta_2=\lambda_1-2\lambda_2+\varpi_2$ around $0^\circ$. From then on, these planets migrate as a pair, while the outer planet is captured in a 1:2 MMR with the second body at about $7\times 10^4$ time units (one of the resonant angles $\theta_3=\lambda_2-2\lambda_3+\varpi_2$ and $\theta_4=\lambda_2-2\lambda_3+\varpi_3$ is in libration). Thus the system is captured in a Laplace-type resonance, whose critical angle is $\phi=\lambda_1-3\lambda_2+2\lambda_3$\footnote{The critical angle is a combination of the resonant angles $\theta_i$, which must respect the d'Alembert characteristics; thus, the sum of all integer coefficients that appear in front of each mean longitude should be zero.}. As the three planets continue to migrate while in resonance, their eccentricities increase. When their values are high enough, the system enters an inclination-type resonance: the angles $\theta_{i_1^2}=2\lambda_1-4\lambda_2+2\Omega_1$, $\theta_{i_2^2}=2\lambda_1-4\lambda_2+2\Omega_2$, $\theta_{i_3^2}=2\lambda_2-4\lambda_3+2\Omega_2$ and $\theta_{i_4^2}=2\lambda_2-4\lambda_3+2\Omega_3$ start to librate around $180^\circ$ at about $3\times 10^5$ time units, and a rapid growth of the inclinations of the three planets is observed.

\begin{figure}
\centering
\rotatebox{270}{\includegraphics[height=12cm]{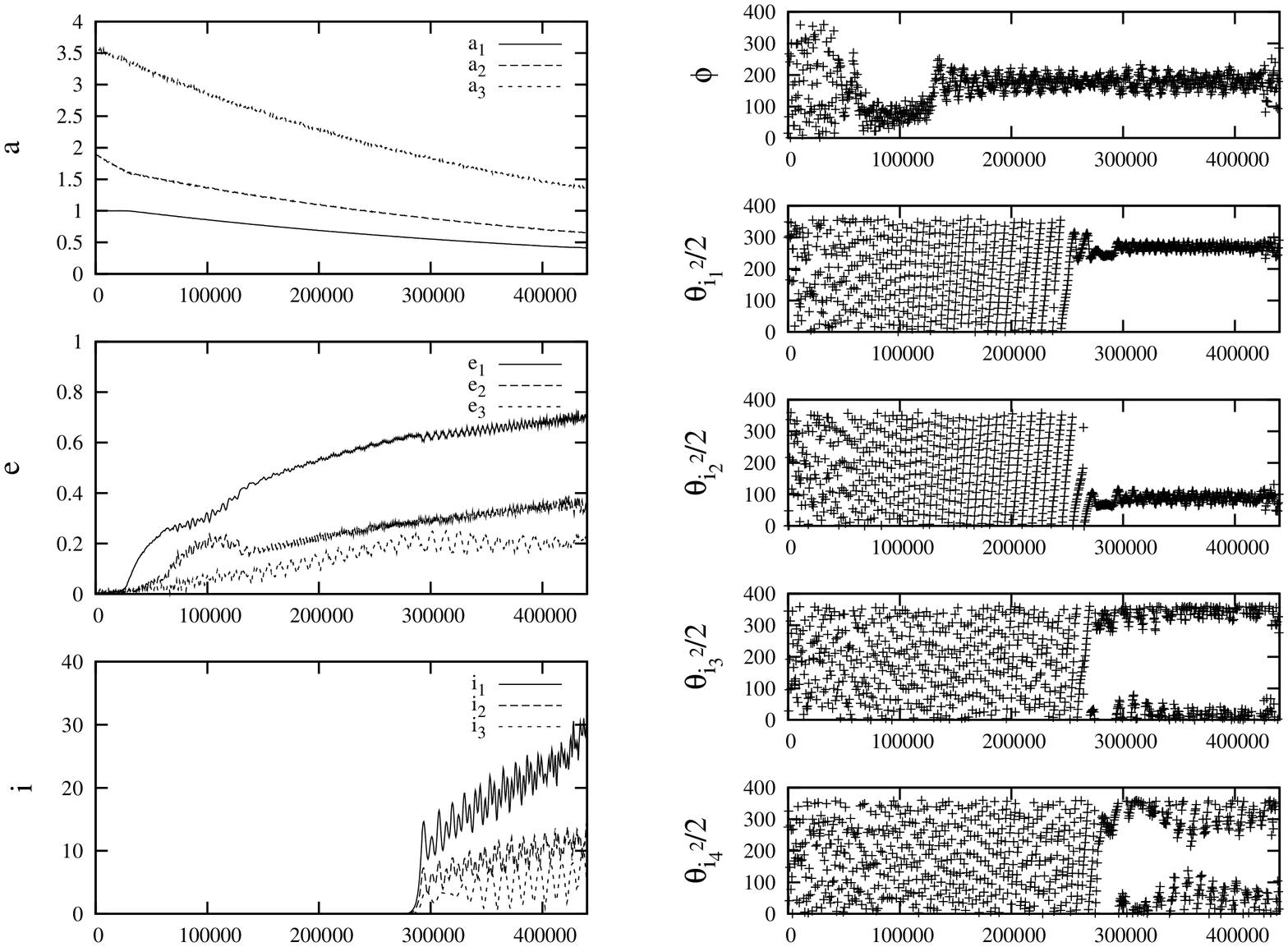}}
\caption{Inclination excitation for a three-planet system in a 1:3:6 resonant configuration. The planetary masses are $m_2=1.5, m_3=0.75 M_{J}$. Initial location of the outer planet is $a_3=3.6$. No eccentricity damping is considered (K=0).}
\label{fig136}
\end{figure} 

\begin{figure}
\centering
\rotatebox{270}{\includegraphics[height=12cm]{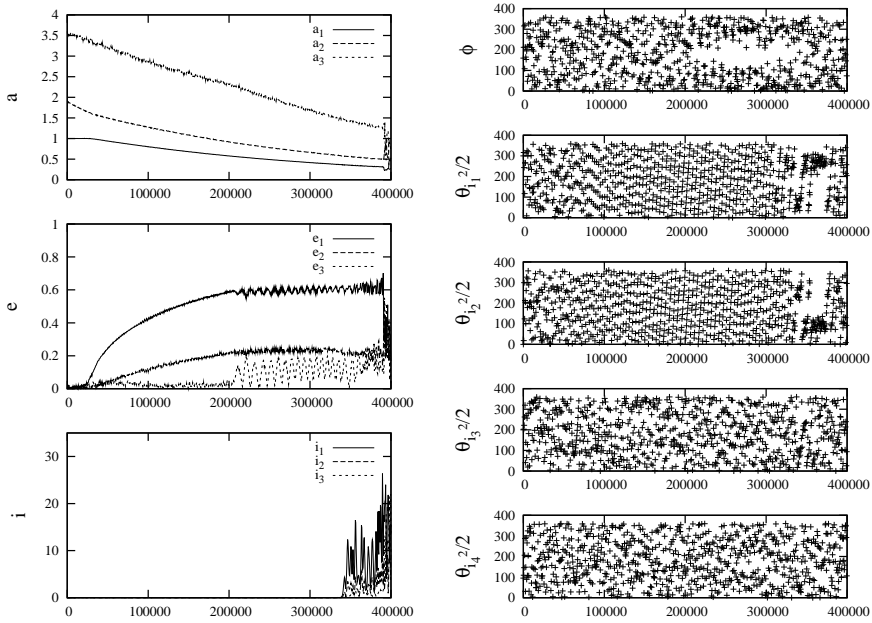}}
\caption{Inclination excitation for a three-planet system in a 1:4:8 resonant configuration. The planetary masses are $m_2=3, m_3=1.5 M_{J}$. Initial location of the outer planet is $a_3=3.6$ and eccentricity damping is set to $K=1$.}
\label{fig148}
\end{figure}

The same type of evolution is observed for higher-order multiple-planet resonances. Figure~\ref{fig136} illustrates the inclination excitation for a three-planet system in a 1:3:6 resonant configuration ($m_1$ and $m_2$ are in a 1:2 MMR, $m_2$ and $m_3$ in a 1:3 MMR), whose resonant angle is $\phi=2\lambda_1-5\lambda_2+3\lambda_3$. The initial configuration of the system is the same as the one of Figure \ref{figthommes} but the planets $m_2$ and $m_3$ are less massive, which favors a 1:3 resonance capture of $m_3$ during its migration towards $m_2$. One of our simulations even ends in a 1:4:8 resonance capture ($m_1$ and $m_2$ are in a 1:2 MMR, $m_2$ and $m_3$ in a 1:4 MMR - see Figure \ref{fig148}), characterized by the libration of $\phi=3\lambda_1-7\lambda_2+4\lambda_3$. However, the inclination excitation mechanism is not effective in this case, as the system is destabilized shortly after the beginning of the libration of the inclination-type resonant angles.

\begin{figure}
\centering{
\rotatebox{270}{\includegraphics[height=12cm]{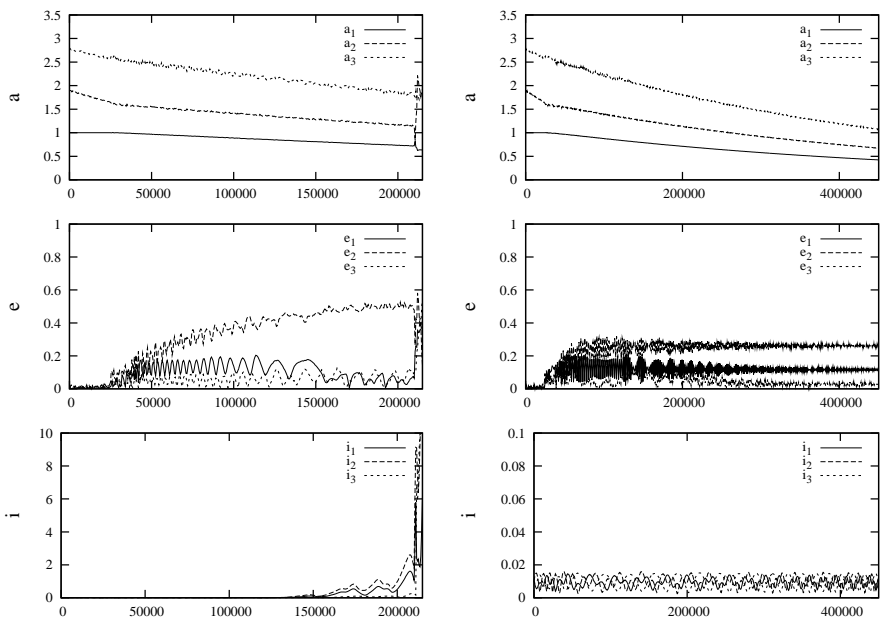}}
}
\caption{Left: Effect of eccentricity damping of the form ${\dot{e}}/{e}=-K\left|{\dot{a}}/{a}\right|$ with $K=1$. The planetary masses are $m_2=0.75, m_3=0.75 M_{J}$. Initial location of the outer planet is $a_3=2.8$. Inclination excitation takes place at $\sim 1.6\times 10^5$ time units, before the disruption of the system at $\sim 2.1\times 10^5$ time units. Right: Same as left Figure, but for $K=5$. Due to the high value of the eccentricity damping, the growth in eccentricity is not sufficient to produce inclination excitation, so that inclinations remain bounded to tiny values (less than $0.02^\circ$).}
\label{figK5}
\end{figure}

Resonant inclination excitation appears to be quite common in our model, as long as eccentricity damping is not too strong. Indeed, a damping rate of $K=5$ (see Equation (\ref{damprate})) does not allow eccentricities to increase sufficiently for an inclination-type resonance to occur, as shown by the example of Figure \ref{figK5}. A similar feature has also been highlighted in the case of a two-planet system in Thommes \& Lissauer (2003) and Libert \& Tsiganis (2009b). Lee \& Thommes (2009) studied the capture of a two-planet system in the 1:2 MMR and found that inclination excitation occurs only for $K\sim 1$ or less, for a range of mass ratios and migration rates similar to what we adopt here. Our simulations confirm this limit and show that it applies also to the case of three-planet resonance\footnote{In that paper another, faster, mode of inclination-excitation was found, not clearly associated with the libration of any critical angle. We have not found an analogue to this mode within our results}.

For no or mild eccentricity damping (i.e.\ $K\leq 1$), Table \ref{tableoutside} suggests that $65\%$ of the triply-resonant systems can go through inclination-type resonance. Hence, significant mutual inclinations between the orbital planes of a three-planet system seems to be a common feature, in the formation scenario depicted by our model.

\begin{figure}
\centering{
\hspace{-6cm}\rotatebox{270}{\includegraphics[height=12cm]{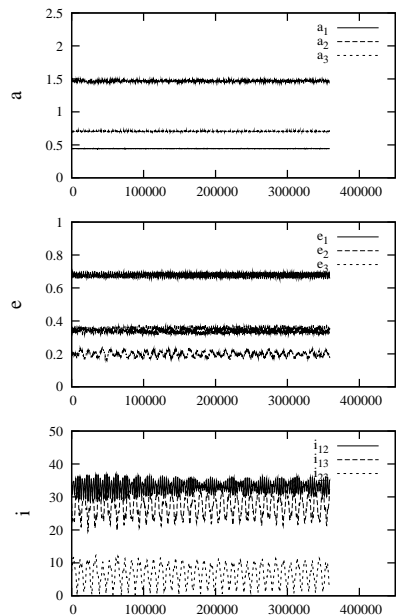}}
}
\caption{Evolution of the system shown in Figure \ref{fig136}, after suddenly switching off the migration force (at $4\times 10^5$ time units). The system remains stable, with high mutual inclinations values between the orbital planes.}
\label{figlgterme}
\end{figure}

Mechanisms responsible for halting Type II migration are not well understood so far. In the case of mild (or no) eccentricity damping, the triply resonant configuration typically becomes unstable, since the eccentricities of the planets keep increasing as migration continues. However, if the disc dissipates before the system dissolves, a stable configuration can be reached. If it occurs during the inclination-type resonance phase, it can produce a stable resonant non-coplanar three-planet system. An example of such a system is given in Figure \ref{figlgterme}, where the migration of the outer planets is abruptly switched off at $4\times 10^5$ time units in the simulation of Figure \ref{fig136} (1:3:6 resonant system). The system remains stable, in orbital resonance (see the constant evolution of the semi-major axes), with small variation in its eccentricities, and with mutual inclinations of the orbital planes between $5^{\circ}$ and $35^{\circ}$. 

An obvious limitation of our model is that we do not account for possible inclination damping by the gaseous disc. Inclined planets would spend most of their orbit away from the disc plane, so it is not easy to find empirical formulae for inclination damping by performing hydrodynamical simulations. On the other hand, if eccentricity damping in discs is moderate (a requirement for the inclination-type resonances to be reached), one should expect the same for inclination damping.

\subsection{Varying the initial eccentricities and migration rates}\label{var}

In the previous section, we assumed planets to be initially on (nearly) circular orbits. The effects of initially eccentric orbits were also studied, in order to check whether capture into multiple-planet resonance can occur for non-circular orbits. The results are given in Table \ref{tableecc}. Three different initial states were analyzed: (A) $e_1=0.15, e_2=0.001, e_3=0.001$, (B) $e_1=0.001, e_2=0.15, e_3=0.001$ and (C) $e_1=0.001, e_2=0.001, e_3=0.15$. For each case, we performed the same set of simulations as in the previous section. As seen from the Table, the main result is that, for no or mild eccentricity damping, the system is easily destabilized before the establishment of any MMR, so much fewer multiply-resonant systems are formed. Most of the inclination excitation observed is due to the 1:2 MMR between the two inner planets. For strong eccentricity damping ($K=5$), the results are much more similar to those presented in the previous section. 

\begin{table}
\centering
\caption{Same as Table \ref{tableoutside}, for different values of the initial eccentricities of the planets. A, B and C denote different initial configurations: (A) $e_1=0.15, e_2=0.001, e_3=0.001$, (B) $e_1=0.001, e_2=0.15, e_3=0.001$ and (C) $e_1=0.001, e_2=0.001, e_3=0.15$. }
\label{tableecc}
\begin{tabular}{llllllllllll}
\hline
& & \multicolumn{3}{r}{$a_3=2.8$} & \multicolumn{3}{r}{$3.1$}& \multicolumn{3}{r}{$3.6$}\\
& $m_2$ & $m_3$ & K=0 & K=1 & K=5 & K=0 & K=1 & K=5 & K=0 & K=1 & K=5 \\
\hline
& 3  & 6 &  &  &  &  &  & 1:2 & & 1:2 & 1:2 $*$ \\
& 3  & 3 &  &  &  & 1:2 &  &  &  &  & \\
& 3  & 1.5 &  &  & 1:2 $*$ & 1:2 & 1:2 & 1:2 &  & 1:2 & \\
& 1.5  & 3 &  &  &  &  & 1:2 &  &  & 1:2 & 1:2 $*$ \\
A & 1.5  & 1.5 &  & 1:2:4  & 1:2:4 &  & 1:2 & 1:2:4 & 1:2 $*$ & 1:2 $*$ & 1:2 $*$\\
& 1.5  & 0.75 &  &  & 1:2:4 & 1:2 & 1:2 & 1:2:4 $*$ & 1:2 &  & 1:2 $*$\\
& 0.75  & 1.5 &  &  & 1:2:4 &  & 1:2 & 1:2:4 & 1:2 &  & 1:2:4 \\
& 0.75  & 0.75 & 1:2:4 $*$ & 1:2:4 & 1:2:4 &  & 1:2 & 1:2:4 &  &  & \\
& 0.75  & 0.375 &  &  & 1:2:4 &  & 1:2 &   &  &  & 1:3:6 \\
\hline
& 3  & 6 &  &  &  &  &  &  &  &  & 1:2  \\
& 3  & 3 &  &  &  &  &  &  & 1:2 $*$ & 1:2  & 1:2 $*$\\
& 3  & 1.5 &  &  &  &  &  & 1:2 $*$ & 1:2  & 1:2 $*$ & 1:2\\
& 1.5  & 3 &  &  &  &  &  &  &  &  & 1:3:6 \\
B & 1.5  & 1.5 &  & 1:2:4 &  &  & 1:2:4 &  &  & 1:3:6 $*$ & \\
& 1.5  & 0.75 &  &  & 1:2:4 &  & 1:2 & 1:2  &  & 1:2 $*$ & \\
& 0.75  & 1.5 &  & 1:2:4 &  &  &  & 1:2 &  &  & 1:2 \\
& 0.75  & 0.75 & 1:2:4 &  & 1:2:4 &  & 1:2 & 1:2:4 &  &  & 1:2\\
& 0.75  & 0.375 & 1:2:4  & 1:2:4  & 1:2:4 &  &  &  1:2 &  &  & 1:2 \\
\hline
& 3  & 6 &  &  &  &  &  &  &  &  & 1:2 $*$ \\
& 3  & 3 &  &  &  &  &  & 1:2 & 1:2 $*$ & 1:2 $*$ & 1:2\\
& 3  & 1.5 &  &  &  & 1:2:4 & 1:2 & 1:2  & 1:2 & 1:2 $*$ & 1:2 $*$\\
& 1.5  & 3 &  &  &  &  &  & 1:2:4 &  &  & \\
C & 1.5  & 1.5 &  &  & &  &  & 1:2 & 1:2 & 1:2 $*$ & 1:3:6\\
& 1.5  & 0.75 &  &  &  & 1:2:5 &  & 1:2 & 1:3:6 $*$ & 1:2 $*$ & 1:3:6\\
& 0.75  & 1.5 &  &  & 1:2:4 & 1:2 & 1:2 & 1:2 & 1:2 &  & 1:3:6 \\
& 0.75  & 0.75 &  &  & 1:2 & 1:2 & 1:2 &  & 1:3:6 $*$ & 1:2 & 1:3:6\\
& 0.75  & 0.375 &  &  & 1:2:4 & 1:2 & 1:2:5 & 1:2:4 & 1:3:6 $*$ & 1:3:6 $*$ & 1:3:6 \\
\hline
\end{tabular}
\end{table}

The new result that can be found in Table \ref{tableecc} is the possibility for planets on initially non-circular orbits of being locked in a 1:2:5 resonance (i.e. $m_1$ and $m_2$ are in a 2:5 MMR, $m_2$ and $m_3$ in a 1:2 MMR). 

\begin{table}
\centering
\caption{Same as Table \ref{tableoutside}, for different values of the migration rate of the planets.}
\label{tablerate}
\begin{tabular}{llllllllllll}
\hline
& & \multicolumn{3}{r}{$a_3=2.8$} & \multicolumn{3}{r}{$3.1$}& \multicolumn{3}{r}{$3.6$}\\
& $m_2$ & $m_3$ & K=0 & K=1 & K=5 & K=0 & K=1 & K=5 & K=0 & K=1 & K=5 \\
\hline
& 3  & 6 &  &  &  &  & 1:2 &  & 1:2 & 1:2 & 1:2 $*$ \\
& 3  & 3 &  &  &  &  & 1:2:4 & 1:2:4  &  & 1:2 & \\
& 3  & 1.5 &  &  & 1:2:4 & 2:3:6 & 1:2 & 1:2:4 & 1:2 $*$ & 1:2 & 1:2 $*$\\
& 1.5  & 3 &  &  &  &  &  & 1:2  &  & &  \\
$\frac{1}{5}\left|\frac{\dot{a}}{a}\right|$ & 1.5 & 1.5 & 1:2  & 1:2:4 & 1:2:4 & 1:2  & 1:2:5 & 1:2:4 & 1:3:6 $*$ & 1:3:6 $*$ & 1:3:6\\
& 1.5  & 0.75 & 1:2:4 & 1:2:4 $*$ & 1:2:4 & 1:2 & 1:2 & 1:2 & 1:3:6 $*$ & 1:3:6 $*$ & 1:3:6\\
& 0.75  & 1.5 & 1:2:4 &  &  & 1:2 &  & 1:2:4 & 1:3:6 & 1:3:6 $*$ & 1:3:6 \\
& 0.75  & 0.75 & 1:2:4 $*$ & 1:2:4 & 1:2:4 &  & 1:2:4 & 1:2 & 1:3:6 $*$& 1:3:6 $*$ & 1:3:6\\
& 0.75  & 0.375 & 1:2:4 $*$ & 1:2:4 $*$ & 1:2:4 & 1:2 & 1:2 & 1:2:4  & 1:3:6 $*$ & 1:3:6 $*$ & 1:3:6 \\
\hline
& 3  & 6 &  &  &  &  &  &  & 1:2 $*$ & 1:2 $*$ & 1:2  \\
& 3  & 3 &  &  &  &  & 1:2 & 1:2  & 1:2 $*$ & 1:2 $*$ & 1:2\\
& 3  & 1.5 &  &  &  & 1:2 $*$ & 1:2 & 1:2 & 1:2 $*$  & 1:2 $*$ & 1:2\\
& 1.5  & 3 &  &  &  &  & 1:2 &  & 1:3:6 & 1:3:6 & 1:3:6 \\
$5\left|\frac{\dot{a}}{a}\right|$ & 1.5  & 1.5 &  &  & 1:2:4 & 1:2 & 1:2:4 & 1:2 & 1:3:6 $*$ & 1:2 $*$ & 1:3:6\\
& 1.5  & 0.75 & 1:2:4 $*$ & 1:2:4 $*$ & 1:2:4 & 1:2 & 1:2:4 &  & 1:3:6 $*$ & 1:3:6 $*$ & 1:3:6\\
& 0.75  & 1.5 &  &  & 1:2:4 &  & 1:2:4 & 1:2 & 1:2 & 1:2 & 1:2 \\
& 0.75  & 0.75 &  & 1:2:4 $*$ & 1:2:4 & 1:2:4 & 1:2:4 & 1:2:4 & 1:2 &  & 1:2\\
& 0.75  & 0.375 & 1:2:4 $*$  &   & 1:2:4 &  & 1:2:4 &  1:2:4 & 1:2 & 1:2 & 2:5:10 \\
\hline
\end{tabular}
\end{table}

\begin{figure}
\centering{
\rotatebox{270}{\includegraphics[height=12cm]{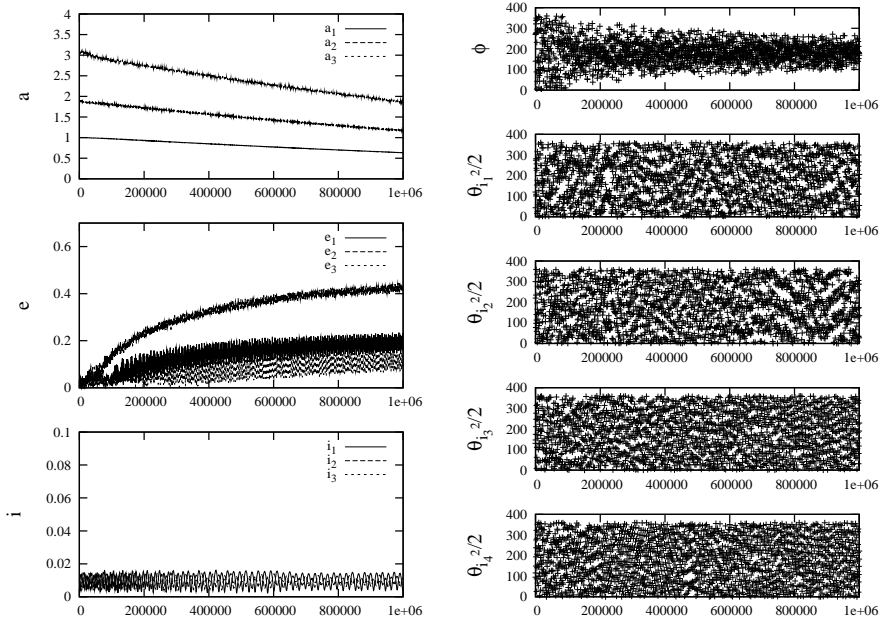}}
}
\caption{Trapping in a 1:2:5 resonance ($m_1$ and $m_2$ are in a 2:5 MMR, $m_2$ and $m_3$ in a 1:2 MMR). The planetary masses are $m_2=1.5, m_3=1.5  M_{J}$. Initial location of the outer planet is $a_3=3.1$ and eccentricity damping is set to $K=1$.}
\label{fig125}
\end{figure} 

The results presented above are by no means a complete study of the dependence of resonance capture on the initial planetary eccentricities. This subject certainly deserves a more thorough analysis. On the other hand, our results clearly show that the probability of capture of three planets in a multiply-resonant, quasi-stable state is much lower, when the initial non-resonant orbits are moderately eccentric and eccentricity damping is not strong.

In the previous section, the migration rates of the outer planets were taken from Equation (\ref{rate}). As explained in that section, this is a rough approximation and therefore the results should be taken with care, as the actual migration mode of the planets may be different. Hence, the influence of the migration rate on the results found in the previous section has to be examined. We analyzed the behavior of the same set of simulations as before, evolving now under five times larger and five times smaller migration rates. As can be seen in Table \ref{tablerate}, the outcomes of the simulations are similar. 
A more careful examination of it shows that actually a slower migration rate favors both the capture in a triple resonance as well as inclination excitation; however the difference is not very dramatic, taking into account that the range of values of $|\dot{a}/a|$ used here corresponds to an order of magnitude. Let us note that one simulation differs again from all the others: the two inner planets are trapped in a 2:5 MMR, before the establishment of the 1:2:5 resonant configuration ($m_1$ and $m_2$ are in a 2:5 MMR, $m_2$ and $m_3$ in a 1:2 MMR, see Figure \ref{fig125}). The resonant angle is $\phi=2\lambda_1-8\lambda_2+6\lambda_3$. The system is disrupted before an inclination-type resonance can be established, due to the large eccentricities reached; apparently the inclination resonance is located for even higher values of $e$ than required for the system to remain stable. The trapping of $m_1$ and $m_2$ in a high order resonance occurs for a slower migration rate; this observation was also highlighted in Libert \& Tsiganis (2009b). 

\subsection{Interior configuration}\label{secondconf}

In the previous simulations, $m_2$ was initially located exterior to its 1:2 MMR with $m_1$. This means that, during its orbital migration, $m_2$ meets (and is usually captured into) this first-order resonance. One may wonder what would happen if $m_1$ and $m_2$ were initially located interior to their 1:2 MMR, or manage to jump across the 1:2 MMR during migration, as is thought to have been the case for the Jupiter-Saturn system (see Pierens \& Nelson 2008). In the following, we set the initial semi-major axis of $m_2$ to $1.4$, and consider three initial values for $a_3$: $2$, $2.4$ and $2.7$. 

\begin{table}
\centering
\caption{Same as Table \ref{tableoutside} for the interior configuration. The migration rate is equal to the value given by Equation (\ref{rate}), namely $\left|\dot{a_2}/a_2\right|=9.03\times10^{-6}$ time units$^{-1}$ and $\left|\dot{a_3}/a_3\right|=5.29\times10^{-6}$, $4.02\times10^{-6}$ or $3.37\times10^{-6}$ time units$^{-1}$ for $a_3=2$, $2.4$ and $2.7$ respectively.}
\label{tableinside}
\begin{tabular}{lllllllllll}
\hline
& \multicolumn{3}{r}{$a_3=2$} & \multicolumn{3}{r}{$2.4$}& \multicolumn{3}{r}{$2.7$}\\
$m_2$ & $m_3$ & K=0 & K=1 & K=5 & K=0 & K=1 & K=5 & K=0 & K=1 & K=5 \\
\hline
3  & 6 &  &  &  &  &  &  &  &  &  \\
3  & 3 &  &  &  &  &  &  &  &  & \\
3  & 1.5 &  &  &  &  &  &  &  &  &\\
1.5  & 3 &  &  &  &  &  &  &  &  &  \\
1.5  & 1.5 & 2:3 & 2:3 &  &  &  &  &  &  & \\
1.5  & 0.75 &  &  &  &  &  &  & &  & \\
0.75  & 1.5 &  &  &  &  &  &  & 2:3 & 3:8:12 $*$ & 4:10:15 \\
0.75  & 0.75 &  &  &  &  &  & 2:3 &  &  &\\
0.75  & 0.375 &  &  &  & 2:3 &  &  &  &  &  \\
\hline
\end{tabular}
\end{table}

\begin{figure}
\centering{
\rotatebox{270}{\includegraphics[height=12cm]{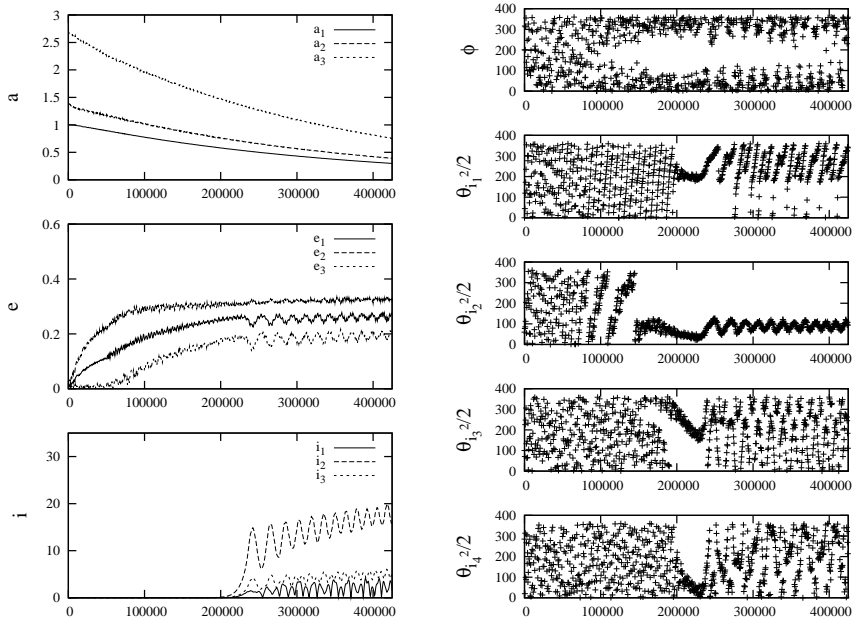}}
}
\caption{Inclination excitation for a three-planet system in a 3:8:12 resonant configuration. The planetary masses are $m_2=0.75, m_3=1.5 M_{J}$. Initial location of the outer planet is $a_3=2.7$ and eccentricity damping is set to $K=1$.}
\label{figinside}
\end{figure} 

For the initial conditions considered above, we have performed the same set of 81 simulations as in Table \ref{tableoutside}. MMR captures are shown in Table \ref{tableinside}. Less than $10\%$ of the simulations result in a two- or three-planet resonance capture. Most of the simulations show a quick destabilization of the system (planet-planet scattering or even mergers). This is most likely an outcome of the initially compact configuration of the system and relatively large assumed masses. In five simulations, $m_1$ and $m_2$ manage to get locked in a 2:3 MMR; this happens when $m_2 \le m_1$. Only two simulations, corresponding to $m_1=1.5$, $m_2=0.75$ and $m_3=1.5 M_{J}$, result in three-planet resonant configurations: a 4:10:15 resonance ($m_1$ and $m_2$ are in a 2:3 MMR, $m_2$ and $m_3$ in a 2:5 MMR), characterized by the libration of $\phi=6\lambda_1-11\lambda_2+5\lambda_3$, and a 3:8:12 resonance ($m_1$ and $m_2$ are in a 2:3 MMR, $m_2$ and $m_3$ in a 3:8 MMR - see Figure \ref{figinside}), characterized by the libration of $\phi=10\lambda_1-18\lambda_2+8\lambda_3$. For this last multiple-planet resonance, an inclination-type resonance is also observed at large-enough eccentricities (see the libration of the resonant angles $\theta_{i_1^2}=4\lambda_1-6\lambda_2+2\Omega_1$, $\theta_{i_2^2}=4\lambda_1-6\lambda_2+2\Omega_2$, $\theta_{i_3^2}=6\lambda_2-16\lambda_3+10\Omega_2$ and $\theta_{i_4^2}=6\lambda_2-16\lambda_3+10\Omega_3$). 

Given the results of this section, it seems that the formation of a three-planet resonant configuration where $m_1$ and $m_2$ are locked in a 2:3 MMR, is rather unlikely, at least for planetary masses larger than $\sim 1M_J$ and mass ratios $m_2/m_1>0.5$. On the other hand, it may be common for mass ratios similar to the Jupiter-Saturn pair. This topic deserves further investigation in the future.

\begin{figure}
\centering
\rotatebox{270}{\includegraphics[height=11cm]{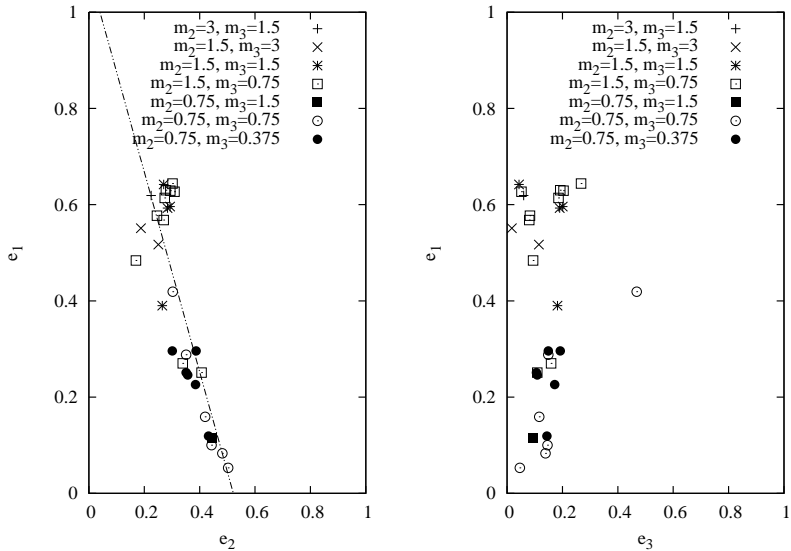}}
\caption{Eccentricity distribution before inclination-type resonance for all systems of Tables \ref{tableoutside} and \ref{tablerate} that lead to inclination excitation}
\label{figecc}
\end{figure} 

\section{Eccentricity distribution}\label{sectionecc}

In this section, we discuss the eccentricity distribution leading to inclination excitation, for the masses, migration rates and initial configurations of the systems presented in Tables \ref{tableoutside} and \ref{tablerate} (i.e.\ initially circular orbits). We plot in Figure \ref{figecc}, the eccentricity values of the three planets, measured just before the increase of the inclinations. All systems of Tables \ref{tableoutside} and \ref{tablerate} that lead to inclination excitation are shown in this figure. We see that inclination excitation occurs for a wide range of eccentricities that depends
strongly on the masses of the two inner planets. For $m_1\le m_2$, the eccentricity of $m_1$ is higher than the one of $m_2$, and vice versa. The eccentricity of the outer body $m_3$ is rather small for all mass ratios considered here. It is interesting to note that left panel of Figure \ref{figecc} shows an apparently correlation between $e_2$ and $e_1$ (see the linear regression). Moreover, no system with eccentricities $e_1$ and $e_2$ both smaller than 0.3 can be found on this plot. This means that inclination-type resonance requires at least one of the two inner planets to reach an eccentricity higher than $e=0.3$.

The eccentricity distribution given in Figure \ref{figecc} is close to the final eccentricities distribution of the resonant systems. Indeed, when the system is in the inclination-type resonance, the eccentricities keep growing but slowly (see e.g.\ Figures 2 and 3), for as long as the system can remain stable.

\section{Conclusion}

In the present work we studied the formation of three-planet resonances in extrasolar planetary 
systems. Such resonances can be established during the phase of gas-driven planet migration. If the resonance persists 
as the gas dissipates, a stable resonant three-planet system can form. Otherwise, the system is dynamically disrupted, 
typically leading to the ejection of one planet. Libert \& Tsiganis (2011) studied the 1:2:4 resonance and showed that, 
if the system becomes unstable, the survivors tend to have large orbital separations (i.e.\ hierarchical systems). 
Of particular interest is the evolution of the orbital inclinations of planets trapped in a 1:2:4 resonance; as migration 
continues and the eccentricities of the planets grow, inclination-type resonances come to play. Thus, similarly to 
the Thommes \& Lissauer mechanism (for two planets), the inclinations can grow to values larger than $\sim 30^{\circ}$.  
  
Here we showed that trapping to several triple resonances can occur, depending of course on the masses of the planets, their initial orbital separations, and the parameters of the disc -- described here in a simplified way through the values of the migration and eccentricity-damping rates. Large planetary masses typically lead to instabilities that destroy the system before a resonance can be established. For masses smaller than $\sim 1.5~M_{J}$, the most common resonant relations found were the $n_3$:$n_2$:$n_1$=1:2:4 
and 1:3:6. These resonances appear to be stable for relatively long times, especially when eccentricity damping is applied. 
For mild eccentricity damping, these resonances can pump the eccentricities of all planets on relatively short time-scales. Then, 
the planets can enter an inclination-type resonance (as in Libert \& Tsiganis 2011), which pumps their inclinations up to $\sim 
35^{\circ}$, thus forming a {\it ``3-D system''}. 

When $m_2$ is started interior to its 1:2 MMR with $m_1$, the strongest resonance in which it could be trapped is the 2:3 MMR with $m_1$. This occurred only in very few cases and, consequently, only two cases of capture in a respective three-planet resonance were found. We note that, according to recent results of hydrodynamical simulations (e.g.\ Pierens \& Nelson 2008), trapping in the 2:3 MMR is considered to be the most likely result of gas-driven migration for the Jupiter-Saturn pair in our solar system. The fact that we do not find such configurations may reflect our large planetary masses (and mass ratios), compared to the Jupiter-Saturn system. Hence, it may actually be the case that trapping in the 2:3 MMR (and related three-planet resonances) is possible only for low-mass planets.  

Our results suggest that trapping in a three-planet resonant configuration during gas-driven migration may be common, as long as the planets in question are not very massive. Also, if eccentricity damping is not too strong in discs, resonant inclination pumping can be efficient, provided that one of the two inner planets eccentricities is higher than 0.3. We note however that our results should be taken with care, as our migration model is not very advanced -- this is why we have avoided presenting statistics on the efficiency of trapping, etc. On the other hand we have no reason to question the generic character of the resonance-trapping and inclination-excitation mechanisms. In order to be able to evaluate the probability of capture in a given triple resonance, its stability time (compared to the disc-evaporation time scale) and its efficiency in exciting the inclinations, we need to resort to more advanced models of migration -- e.g.\ the one used in Thommes et al.\ (2008) -- that take into account variations in the gas surface density induced by the relative motion of the planets, as well as gas dissipation. We hope to be able to report on these issues in a forthcoming paper.

\begin{acknowledgements}
The work of A.-S. Libert is supported by an FNRS Postdoctoral Research Fellowship. Numerical simulations were made on the local computing resources (Cluster URBM-SYSDYN) at the University of Namur (FUNDP, Belgium). 

\end{acknowledgements}



\begin{thebibliography}{}
%

%



\makeatletter 
\renewcommand{\@biblabel}[1]{} 
\makeatother 

\bibitem{Ada03}
Adams, F.C., Laughlin, G.: Migration and dynamical relaxation in crowded systems of giant planets. Icarus 163, 290-306 (2003)

\bibitem{Beau05}
Beaug\'e, C., Michtchenko, T.A., Ferraz-Mello, S.: Planetary migration and extrasolar planets in the 2/1 mean-motion resonance. MNRAS 365, 1160-1170 (2005)

\bibitem{Beau08}
Beaug\'e, C., Giuppone, C.A., Ferraz-Mello, S., Michtchenko, T.A.: Reliability of orbital fits for resonant extrasolar planetary systems: the case of HD82943. MNRAS, 385, 2151-2160 (2008)

\bibitem{Cha08}
Chatterjee, S., Ford, E.B., Matsumura, S., Rasio, F.A.: Dynamical Outcomes of Planet-Planet Scattering. ApJ 686, 580-602 (2008)

\bibitem{Dun98}
Duncan, M.J., Levison, H.F., Lee, M.H.: A Multiple Time Step Symplectic Algorithm for Integrating Close Encounters. Astron. J. 116, 2067-2077 (1998) 

\bibitem{Fab07}
Fabrycky, D., Tremaine, S.: Shrinking Binary and Planetary Orbits by Kozai Cycles with Tidal Friction. ApJ 669, 1298-1315 (2007)

\bibitem{FM03}
Ferraz-Mello, S., Beaug\'e, C., Michtchenko, T.A.: Evolution of Migrating Planet Pairs in Resonance. Celest. Mech. \& Dyn. Astro. 87, 99-112 (2003)

\bibitem{For05}
Ford, E.B., Lystad, V., Rasio, F.A.: Planet-planet scattering in the upsilon Andromedae system. Nature 434, 873-876 (2005)

\bibitem{For08}
Ford, E.B., Rasio, F.A.: Origins of Eccentric Extrasolar Planets: Testing the Planet-Planet Scattering Model. ApJ 686, 621-636 (2008)

\bibitem{go80}
Goldreich, P., Tremaine, S.: Disk-satellite interactions. ApJ 241, 425-441 (1980) 

\bibitem{had06}
Hadjidemetriou, J.D.: Symmetric and asymmetric librations in extrasolar planetary systems: a global view. Celest. Mech. \& Dyn. Astro. 95, 225-244 (2006)

\bibitem{had09}
Hadjidemetriou, J.D., Psychoyos, D., Voyatzis, G.: The 1/1 resonance in extrasolar planetary systems. Celest. Mech. \& Dyn. Astro. 104, 23-38 (2009)

\bibitem{Jur08}
Juri\'c, M., Tremaine, S.: Dynamical Origin of Extrasolar Planet Eccentricity Distribution. ApJ 686, 603-620 (2008)

\bibitem{Lee02}
Lee, M.H., Peale S.J.: Dynamics and Origin of the 2:1 Orbital Resonances of the GJ 876 Planets. ApJ 567, 596-609 (2002)

\bibitem{Lee09}
Lee, M.H., Thommes, E.W.: Planetary Migration and Eccentricity and Inclination Resonances in Extrasolar Planetary Systems. ApJ 702, 1662-1672 (2009)

\bibitem{Las97}
Laskar, J.: Large scale chaos and the spacing of the inner planets. A\&A, 317, L75-L78 (1997)

\bibitem{Lib07}
Libert, A.-S., Henrard, J.: Exoplanetary systems: The role of an equilibrium at high mutual inclination in shaping the global behavior of the 3-D secular planetary three-body problem. Icarus 191, 469-485 (2007)

\bibitem{Lib08}
Libert, A.-S., Henrard, J.: Secular frequencies of 3-D exoplanetary systems. Celest. Mech. \& Dyn. Astro. 100, 209-229 (2008)

\bibitem{Lib09a}
Libert, A.-S., Tsiganis, K.: Kozai resonance in extrasolar systems. A\&A 493, 677-686 (2009a)

\bibitem{Lib09b}
Libert, A.-S., Tsiganis K.: Trapping in high-order orbital resonances and inclination excitation in extrasolar systems. MNRAS 400, 1373-1382 (2009b)

\bibitem{Lib11}
Libert, A.-S., Tsiganis K.: Formation of '3D' multiplanet systems by dynamical disruption of multiple-resonance configurations. MNRAS 412, 2353-2360 (2011)

\bibitem{Lis11}
Lissauer, J.J., Ragozzine, D., Fabrycky, D.C., Steffen, J.H., Ford, E.B. et al.: Architecture and Dynamics of Kepler's Candidate Multiple Transiting Planet Systems. arXiv:1102.0543, submitted to ApJ (2011)

\bibitem{Mar02}
Marzari, F., Weidenschilling, S.J.: Eccentric Extrasolar Planets: The Jumping Jupiter Model. Icarus 156, 570-579 (2002)

\bibitem{Mat10}
Matsumura, S., Thommes, E.W., Chatterjee, S., Rasio, F.A.: Unstable Planetary Systems Emerging Out of Gas Disks. ApJ 714, 194-206 (2010)

\bibitem{Mc10}
McArthur, B.E., Benedict, G.F., Barnes, R., Martioli, E., Korzennik, S., Nelan, Ed., Butler, R.P.: New Observational Constraints on the υ Andromedae System with Data from the Hubble Space Telescope and Hobby-Eberly Telescope. ApJ 715, 1203-1220 (2010)

\bibitem{Mic06a}
Michtchenko, T.A., Beaug\'e, C., Ferraz-Mello, S.: Stationary Orbits in Resonant Extrasolar Planetary Systems. Celest. Mech. \& Dyn. Astro. 94, 411-432 (2006a)

\bibitem{Mic06b}
Michtchenko, T.A., Ferraz-Mello, S., Beaug\'e, C.: Modeling the 3-D secular planetary three-body problem. Discussion on the outer $\upsilon$ Andromedae planetary system. Icarus 181, 555-571 (2006b)

\bibitem{Moo05}
Moorhead, A.V., Adams, F.C.: Giant planet migration through the action of disk torques and planet planet scattering. Icarus 178, 517-539 (2005)

\bibitem{Mor07b}
Morbidelli, A., Crida, A.: The dynamics of Jupiter and Saturn in the gaseous protoplanetary disk. Icarus 191, 158-171 (2007)

\bibitem{Mor07}
Morbidelli, A., Tsiganis, K., Crida, A., Levison, H.F., Gomes, R.: Dynamics of the Giant Planets of the Solar System in the Gaseous Protoplanetary Disk and Their Relationship to the Current Orbital Architecture. AJ 134, 1790-1798 (2007)

\bibitem{Nag08}
Nagasawa M., Ida S., Bessho T.: Formation of Hot Planets by a Combination of Planet Scattering, Tidal Circularization, and the Kozai Mechanism. ApJ 678, 498-508 (2008)

\bibitem{pa01}
Papaloizou, J.~C.~B., Nelson, R.~P., Masset, F.: Orbital eccentricity growth through disc-companion tidal interaction. A\&A 366, 263-275 (2001) 

\bibitem{pie08}
Pierens, A., Nelson, R.~P.: Constraints on resonant-trapping for two planets embedded in a protoplanetary disc. A\&A 482, 333-340 (2008) 

\bibitem{rei09}
Reidemeister, M., Krivov, A.V., Schmidt, T.O.B. et al.: A possible architecture of the planetary system HR 8799. A\&A 503, 247-258 (2009)

\bibitem{riv10}
Rivera, E.J., Laughlin, G., Butler, R.P., Vogt, S.S., Haghighipour, N., Meschiari, S.: The Lick-Carnegie Exoplanet Survey: a Uranus-Mass Fourth Planet for GJ 876 in an Extrasolar Laplace Configuration. ApJ 719, 890-899 (2010)

\bibitem{sha73}
Shakura, N.I., Sunyaev, R.A.: Black holes in binary systems. Observational appearance. A\&A 24, 337-355 (1973)

\bibitem{Tho03}
Thommes, E.W., Lissauer, J.J.: Resonant Inclination Excitation of Migrating Giant Planets. ApJ 597, 566-580 (2003)

\bibitem{Tho08}
Thommes, E.W., Matsumura, S., Rasio, F.A.: Gas Disks to Gas Giants: Simulating the Birth of Planetary Systems. Science 321, 814-817 (2008)

\bibitem{wa97}
Ward, W.~R.: Protoplanet Migration by Nebula Tides. Icarus 126, 261-281 (1997) 


\end{thebibliography}
\end{document}